\documentclass[11pt]{amsart}
\usepackage{geometry}                
\usepackage[british]{babel}
\usepackage{graphicx}
\usepackage{amssymb}
\usepackage{epstopdf}
\usepackage[lined,algonl,boxed]{algorithm2e}
\usepackage{alltt}
\usepackage{float}

\newtheorem{prop}{Proposition}

\newcommand{\latset}{$\mathcal I(S_H)$\ }

\floatstyle{boxed}
\newfloat{program}{thp}{lop}
\floatname{program}{Program}

\begin{document}
\title{Quantifying Information Leak Vulnerabilities}
\author{Jonathan Heusser${}^{\star}$}

\author{Pasquale Malacaria${}^{\dagger}$}
\date{6 July 2010, $\star$ \texttt{jonathan.heusser@dcs.qmul.ac.uk} $\dagger$ \texttt{pm@dcs.qmul.ac.uk}}

\begin{abstract}
Leakage of confidential information represents a serious security risk.
Despite a number of novel, theoretical advances, it has been unclear if and how quantitative approaches to measuring leakage of confidential information could be applied to substantial, real-world programs. This is mostly due to the high complexity of computing precise leakage quantities.
In this paper, we introduce a technique which makes it possible to decide if a program conforms to a quantitative policy which scales to large state-spaces with the help of bounded model checking.

Our technique is applied to a number of officially reported information leak vulnerabilities in the Linux Kernel. Additionally, we also analysed authentication routines in the Secure Remote Password suite and of a Internet Message Support Protocol implementation. Our technique shows when there is unacceptable leakage; the same technique is also used to verify, for the first time, that the applied software patches indeed plug the information leaks.

This is the first demonstration of quantitative information flow addressing security concerns of real-world industrial programs.
\end{abstract}
\maketitle

\section{Introduction}
Quantitative Information Flow (QIF) \cite{0dav, POPL07} aims to provide techniques and tools able to quantify leakage of confidential information. As a motivating example
consider a prototypical password checking program

\begin{quote}
\begin{alltt}
if (password==guess) access=1; else access=0;
\end{alltt}
\end{quote}
Notice how there is an unavoidable leakage of confidential information in this program:
 an attacker observing the value of {\tt access} will be able to infer if he guessed the right password (complete leakage if he did guess it right) and if the guess was wrong he will have eliminated one possibility from the search space. Notice also how essential the {\em amount of information leaked} is: if the amount leaked is very small then the program could as well be considered secure.

If, as the above example illustrates, leakage is somehow unavoidable then the real question is not whether or not programs leak, but how much. This point is what makes Quantitative Information Flow  an appealing theory.
In a nutshell, QIF aims to measure the amount of information from confidential data (in the above example the variable {\tt password}) that an attacker who can read/write the public input data ({\tt guess}) will be able to infer from some observable variable ({\tt access}).

However, implementing a precise QIF analysis for secret sizes of more than a few bits is computationally infeasible;
roughly speaking this is because classical QIF computes the entropy of a random variable whose complexity is the same as computing all possible runs of the program. Even when abstraction techniques and statistical sampling are integrated with QIF  \cite{KoRy10}  to help the scalability issues  a useful analysis for real code still seems problematic.

In this paper, we introduce a useful quantitative analysis for C code: we will demonstrate the analysis on reported  information leakage vulnerabilities  in the Linux Kernel and the PHP interpreter. All of the covered vulnerabilities are referenced by the standardised vulnerability repository CVE from Mitre\footnote{http://cve.mitre.org, CVE is industry-endorsed with over 70 companies actively involved.
}. 
%

To address the computational feasibility of the quantitative analysis we shift the focus from the question ``How much does it leak?'' to the simpler quantitative question ``Does it leak more than k?''. We will show how the questions are related and more importantly we will show that off-the-shelf symbolic model checkers like CBMC \cite{ckl2004} are able to efficiently answer the second kind of question.
CBMC is a good choice for several reasons: (i) it makes it easy to parse and analyse large ANSI-C based projects (ii) it models bit-vector semantics of C accurately which makes it able to detect arithmetic overflows amongst others, which turns out to be important (iii) nondeterministic choice functions can be used to easily model user input, which also enjoys efficient solving due to the symbolic nature of the model checker (iv) despite being a bounded model checker, CBMC can check whether enough unwinding of the transition system were performed to prove that there are no deeper counterexamples.

Our experiments show that the analysis not only quantifies the leakage but also helps in understanding the nature of the leak.
In particular, the counterexample produced by the model checker, when a leakage property is violated, can provide insights into the cause of the leak. For example, we can extract a public user input from the counterexample needed to trigger a violation.

Another surprising result of our experiment is that in certain circumstances we were able to use our technique to prove whether the official patch provided for the vulnerability does actually eliminate the information leak. This is achieved by point (iv) from above, when the model checking process is actually complete.

In summary the main technical contributions of this paper are the following:
\begin{enumerate}
\item We present the first quantitative leakage analysis of systems software.
\item We show how to express Quantitative Information Flow properties  that can be efficiently checked using bounded symbolic model checking.
\item We show that the technique not only quantifies leakage in real code but also provides valuable information about the nature of the leak.
\item In some cases we are able to prove that official patches for reported
vulnerability do indeed eliminate leakage; these constitute the first positive proofs of absence of QIF vulnerabilities for real-world systems programs.
\end{enumerate}


\section{Model of Programs and Distinctions}

We aim to model the input/output behaviour of a C function where inputs are formal arguments to the function and outputs are either return values or pointer arguments.

In the following we will consider $P$ to be  a C function taking high and low inputs noted $h,l$; 
we call observables low variables whose values are ``publicly available'' after running $P$.
As an example consider the following ``modulo'' program
\begin{quote}
\begin{verbatim}
o = (h % 4 ) + l  
\end{verbatim}
\end{quote}
and suppose {\tt h} is a 4 bits variable with values 0..15 and {\tt l} a 1 bit variable with values 0,1; then the low input for $P$ is the variable $l$  and the observable is the variable $o$ whose possible values are 0..5. 

Formally, a program $P$ is modelled as transition system $TS = (S, T, I, F)$ with $S$ being the program states, $T \subseteq S \times S$ are the program transitions and $I$ the initial states and $F$ the final states. Let us define a successor function for a state $s\in S$
\begin{equation*}
\text{Post}(s) = \{s' \in S\ \vert\ (s,s') \in T\}
\end{equation*} 
A state $s$ is in $F$ if $\text{Post}(s) = \emptyset$. A path is a finite sequence of states $\pi = s_{0} s_{1} s_{2} \dots s_{n}$ such that $s_{0} \in I$ and $s_{n} \in F$.

A state is a tuple $S = S_{H} \times S_{L}$ of the pair of confidential input $H$ and low input $L$. We consider initial/final or input/output pairs of states of a path,  $\langle (h, l), o \rangle$ where the second component is the output $o$ produced by the final state drawn from some output alphabet $O$. In the above example an input/output pair would be $\langle (5, 1), 2 \rangle$ representing the computation
$ (5\% 4)+1=2$.


Confidential inputs are denoted as $h \in H$, low inputs $l \in L$, and low observations $o \in O$, where the output behaviour of the function is always a low observation and the input is an initial state $(h,l)$.
A distinction on the confidential input through observations $O$ is one where there exists at least two paths through $P$, modelled as $TS$, which leads to different observations for different confidential input but constant low input. 

We define an equivalence relation $\simeq_{P,l}$ on the values of the high variables as follows: 
$h \simeq_{P,l} h' $ iff if   $\langle (h, l), o \rangle$, $\langle (h', l), o' \rangle$  are  input/output pairs in $P$ then $o=o'$.

Hence, two high values are equivalent (w.r.t. a low value $l$) if they cannot be distinguished by any observable. In the running example an equivalence class in $ \simeq_{P,1}$ would for example be $\{ 1,5,9,13  \}$.
The equivalence relation associated to $P, l$ is an element of the set of all possible equivalence relation on the values of high. 

Let \latset as the set of all possible equivalence relations on a set $S_H$. Define on \latset the order:
\begin{equation}\label{ORDER}
\approx\ \sqsubseteq\ \sim\ \leftrightarrow\ \forall s_1, s_2\ ( s_1 \sim s_2 \Rightarrow s_1 \approx s_2)
\end{equation}
where $\approx,\sim$  $\in $ \latset and $s_{1},s_{2} \in S_H$. 
$\sqsubseteq$ defines a complete lattice over $S_H$. It is a {\em refinement order} with bottom element being the relation relating every state and top element being the identity relation.
This is described as the Lattice of Information \cite{loi}.

Non leaking programs (i.e. satisfying  non-interference \cite{GoMe}) are characterised as follows:
\begin{prop}
$ P$ is non-interfering iff for all $l$, $\simeq_{P,l}$ is the least element in \latset.
\end{prop}

An attacker controlling the low inputs can be modelled by an equivalence relation  $\simeq_{P}$ corresponding to a particular $\simeq_{P,l}$. 
 
Formally, we define a {\em quantitative policy} as a non-negative natural number $N$. A relation $\simeq_{P,l}$ breaches a policy if  $|\simeq_{P,l}|>N$  (where $| \simeq_{P,l}|$ is the number of equivalence classes of $\simeq_{P,l}$).
 In our model, an attacker will always choose a relation breaching the policy, provided that given a policy and a program such a relation exists. We use  $\simeq_{P}$ with the program $P$ being initialised with the attacker's choice of $l$\footnote{In the paper such attacker choices will be modelled by the nondeterministic choice function {\tt input()}.}.
 
 In the above example, a choice could be $\simeq_{P}\ =\ \simeq_{P,0}$ corresponding to the program {\tt l=0; o = (h \% 4 ) + l}.

Quantitative Information Flow uses information theoretical measures like Shannon entropy to measure leakage of confidential information. The measure of a program can be broken down into two main steps \cite{POPL07,aqua}: 
\begin{enumerate}
\item interpret the program as a random variable $R_P$
\item compute the entropy of $R_P$ (noted $H(R_P)$)
\end{enumerate}
It has been shown that  $R_P$ and $\simeq_P$ coincide \cite{POPL07,Bert}.
For example for the modulo program above under the assumption of uniform distribution on the input there are 4 equivalence classes each having probability $\frac{1}{4}$. The Shannon entropy of that program is then 
\[4*- \frac{1}{4} \log_2(\frac{1}{4})=  2 \]
This number 2 represents the fact that the observations reveal which of the 4 possible classes (i.e. 2 bits of information) the high input belongs to.

 $R_P$ and $\simeq_P$  are also order related as the following proposition shows \cite{aqua}:
 \begin{prop}\label{LoIH}
 $ \simeq_P\ \sqsubseteq\ \simeq_{P'}$ iff for all probability distributions $H(R_P)\leq H(R_{P'})$
 \end{prop}
 
 To further understand the importance of $ \simeq_P$ in Quantitative Information Flow we need to introduce the information theoretical concept of channel capacity: consider the password check example from the introduction. Suppose the password is a 64 bits randomly chosen string; we have two equivalence classes, one with 1 element so having probability $\frac{1}{2^{64}}$, the other class with $2^{64}-1$ elements having thus probability $1-\frac{1}{2^{64}}$. The entropy is then  $3.46944695 \times 10^{-18}$: as expected a password check of a big password should leak very little. Suppose however that the probabilities of the high inputs are such  that both equivalence classes have probability $\frac{1}{2}$. Then the entropy dramatically raises to 1 which is the channel capacity, i.e. the maximum leakage achievable given two classes: $\log_2(2)=\log_2(|\simeq_P|)$.
In the modulo example the channel capacity is 2 which happens to be given by the uniform distribution on the high input. Other distributions on the high input cannot give higher entropy: for example if we consider the distribution where all even numbers have equal probability $\frac{1}{8}$, and all odd numbers have 0 probability then the resulting entropy will be 1.

The following result establishes basic relationships between leakage, channel capacity, and number of distinctions:
\begin{prop}\label{chancap}\mbox{}
\begin{enumerate}
\item  $P$ is non-interfering iff $\log_2(| \simeq_P |)=0$
\item The channel capacity\footnote{The channel capacity is the maximum possible leakage where we consider all possible probability distributions on the inputs \cite{PLAS08}} of $P$ is $\log_2(| \simeq_P|)$ .
\item If for all probability distributions $H(R_P)\leq H(R_{P'})$ then $| \simeq_P| \leq | \simeq_{P'}|$
\end{enumerate}
\end{prop}
Point (1) is proved in \cite{CHM05}, (2) in \cite{PLAS08} and (3) is a consequence of proposition \ref{LoIH} whose proof is in \cite{aqua}.
Hence a lower bound on $| \simeq_P |$ provides a lower bound on the channel capacity of the program $P$.

Hence, because of proposition \ref{chancap} the inequality $| \simeq_P |> N$, which is at the heart of our analysis, can be rephrased to the following statement: in a setting where the distribution of the secret is the most favourable for the attacker then the leakage is at least $\log_2(N)$ bits.

\section{Encoding Distinction-Based Quant Policies}

Recall that for a program $P$ a quantitative policy is a natural number $N$ which limits the cardinality of $\simeq_{P}$ to $N$.

In other words, a program violates a quantitative policy if it makes more distinctions than what is allowed in the policy. A leaking programs is one breaching the policy $N=1$ in the above definition.

We take ideas from assume-guarantee reasoning \cite{assumeguarantee} to encode such a policy in a driver function, which tries to trigger a violation, i.e. producing a counterexample, of the policy. If the policy states that the function \texttt{func} is not allowed to make more than 2 distinctions then this is modelled as shown in Program \ref{PROG_1}. This driver only has a high component as a state, which is passed to the function \texttt{func} where the policy is tested on.

\begin{program}
\begin{small}
\begin{alltt}
int h1,h2,h3;
int o1,o2,o3;

h1 = input();  h2 = input(); h3 = input();

o1 = func(h1);
o2 = func(h2);
assume(o1 != o2); // (A)

o3 = func(h3);
assert(o3 == o1 || o3 == o2); // (B)
\end{alltt}
\end{small}
\caption{Example driver checking for 2 distinctions}
\label{PROG_1}
\end{program}

Drivers always have a similar structure: we model the secret by a nondeterministic choice function \texttt{input()} as a placeholder for all possible values of that type; then for a policy of checking for $N$ distinctions, the function under inspection is called $N$ times. The crucial step (A) is the use of the \texttt{assume} statement after the calls: the driver assumes that, in this case, there are two different return values found already. The function is called an $N+1$th time and at (B) the driver asserts that the next output is either one of the previously found outputs. 

The \texttt{assume} statement only considers execution paths which satisfy the given boolean formula, all other paths are rejected. Further, the bounded model checker used will try to find a counterexample to the negated assertion claim, which is only satisfiable if and only if a counterexample exists. If it is unsatisfiable means that the original claim holds, i.e. the program conforms to the policy. The verification condition generated by the bounded model checker for the policy in Program \ref{PROG_1} is: 
\begin{quote}\begin{alltt}o1 != o2 \(\implies\) (o3 == o1 || o3 == o2)\end{alltt}\end{quote}
Where the bounded model checker tries to find a counterexample (execution path) using the negated claim such that the following holds
\begin{quote}\begin{alltt}o1 != o2 \(\land\) o3 != o1 \(\land\) o3 != o2\end{alltt}\end{quote}
i.e. that there are three distinctions possible.

\begin{figure}[t]
\begin{small}
\begin{algorithm}[H]
\dontprintsemicolon
\SetLine
\KwIn{Function \texttt{func}, types \texttt{t,t',t''}, comparison \texttt{eq\_t}, bound $k$, threshold $N$}
\KwOut{Driver.c}
\begin{alltt}
t o_1, \(\dots\), o_n, o_{n+1};
t' h_1, \(\dots\), h_n, h_{n+1};
t'' l;

h_1 = input(); \(\dots\) h_n = input();
l = input();
o_1 = func(h_1, l);
\(\vdots\)
o_{n} = func(h_{n}, l);
assume(!eq_t(o_1, o_2) && !eq_t(o_1, o_3) && \(\dots\));

o_{n+1} = func(h_{n+1}, l);
assert(eq_t(o_{n+1}, o_1) || eq_t(o_{n+1}, o_2) || \(\dots\));
\end{alltt}
\caption{Template to syntactically generate a driver for an $N$ distinction policy}
\label{ALGO_1}
\end{algorithm} 
\end{small}
\end{figure}

Another possibility is that the function \texttt{func} does not even make two distinctions, such that the \texttt{assume} statement  at point (A) is always false, which leads to proving the policy (or any policy) vacuously true, because for any assertion \texttt{Q} the verification condition is true, i.e. $\texttt{false} \implies \texttt{Q}$.

\subsection{Bounded Model Checking}

We use the bounded model checker {\sc CBMC} to verify or falsify a policy. CBMC encodes an ANSI-C program into a propositional formula by unwinding the transition relation and user defined specifications up to some bound. This formula is only satisfiable if there exists an error trace violating the specification. 

The tool can also check if the unwinding bound is sufficient by introducing \textit{unwinding assertions}, which are assertions on the negated loop guards. This ensures that no longer counterexample can exist than the used bound. To \textit{prove} any properties the analysis has to pass unwinding assertions, otherwise it can only be used as a way to find counterexamples up to the unwinding bound.

The C program gets encoded into constraints $C$ and the property -- user defined assertions -- are encoded in $P$.
Then the model checker tries to find a satisfiable assignment to the formula
\begin{equation*}
C \land \lnot P
\end{equation*}
where $P$ is an accumulation of the assumptions and assertions made in the program text. Thus if there are two \texttt{assume} statements in the driver  with expressions $E_{1}$ and $E_{2}$ and one \texttt{assert} statement with expression $Q$ then $P$ is
\begin{equation*}
P \equiv E_{1} \land E_{2} \implies Q
\end{equation*}

\subsection{Driver}
A general template for a driver is described in Algorithm \ref{ALGO_1}. The inputs to the algorithm are the function \texttt{func} to be analysed, possibly up to three different types for the input/output pair $\langle (h, l), o \rangle$, and a comparison function \texttt{eq\_t} which returns true if the arguments of type \texttt{t} are equal, where  \texttt{t} is the type of the observation of function \texttt{func}. This comparison function could be as simple as \texttt{==} of C, or a more complex function, such as \texttt{memcmp}, if \texttt{t} is an array or string. Also note, that the observations \texttt{o\_i} do not need to be only return values, but can also be pointer arguments to \texttt{func}.

\begin{prop}[Correctness of driver template]
If the driver template in Algorithm \ref{ALGO_1} is successfully verified up to a bound $k$ (i.e. the negated claim is unsatisfiable) then the function \texttt{func} does not make more than $N$ distinctions on the output within the bound $k$.
Formally, we state that the validity of the driver implies the validity of the following implication
\begin{equation*}
o_{1} \neq o_{2} \land o_{1} \neq o_{3} \land \dots \land o_{n-1} \neq o_{n} \implies o_{n+1} = o_{1} \lor \dots \lor o_{n+1} = o_{n}
\end{equation*}
\end{prop}

Thus, we can make the following claims on the result of the model checking process: For a given bound $k$ and a policy,
\begin{itemize}
\item if the model checker finds a counterexample then the policy is violated, i.e. the program makes more distinctions than specified


\item if the process ends with a successful verification of the policy without unwinding assertions then the policy holds up to an unwinding of $k$.

\item if the process ends with a successful verification of the policy \textit{with} unwinding assertions then the policy holds for any number of iterations.
\end{itemize}


\section{Checking Quantitative Policies}

The steps in checking a program or function for the compliance with a quantitative policy are as follows:
(1) Define the input state $(h,l)$ and output state $o$ in the code, i.e. the confidential input $h$, the low input $l$ and the observation $o$
(2) Define the maximum number of distinctions in the policy and an unwinding factor $k$
(3) Generate a driver function using the template in Algorithm \ref{ALGO_1}
(4) Run {\sc CBMC} on the driver. If the driver is successfully verified, potentially increase the unwinding factor.

\subsection{Modelling Low Input}\label{SEC:MODELLING}
A crucial aspect of  our analysis is to model low user input, which is most of the time responsible for triggering a bug which causes the information leak. These bugs  only happen on a very restricted number of execution paths and could be exploited by a malicious user choosing a special user input. This scenario generally applies when studying many CVE reported information leakage vulnerabilities.

Let us look at the following simplified code in Program \ref{PROG_2}, which contains an integer underflow, taken from the vulnerability CVE-2007-2875 in the linux kernel.

\begin{program}
\begin{small}
\begin{verbatim}
typedef long long loff_t;
typedef unsigned int size_t;
int underflow(int h, loff_t ppos) {
  int bufsz;
  size_t nbytes;
  bufsz=1024;
  nbytes=20;
  
  if (ppos + nbytes > bufsz) // (A)
       nbytes = bufsz - ppos; // (B)
  if(ppos + nbytes > bufsz) {
     return h; // (C)
  } else {
     return 0;
  }
}
\end{verbatim} 
\end{small}
\caption{Integer underflow causing a leak}
\label{PROG_2}
\end{program}
At first, it seems not possible that the point (C) where the secret \texttt{h} gets returned is ever executed, because exactly that check is done in (A) which reduces the variable \texttt{nbytes} to be within the bound \texttt{bufsz}. However, due to wrong choice and combination of types, the subtraction in (B) causes an underflow in \texttt{nbytes} for a very large \texttt{ppos} value. And unfortunately, \texttt{ppos} is a user controlled input variable, such that when its value is chosen correctly, point (C) is reached. 

In this case, a state in the system is the tuple $(h, l)$ which represents the arguments to the function \texttt{underflow}, i.e. the formal parameters \texttt{h} and \texttt{ppos}; observations are the return values of this function. 
The generated driver can automatically find the low part of a state which triggers such subsequent information leaks, because the analysis instructs the model checker to find \textit{any} possible execution path satisfying the assumptions and assertions on the outputs, given nondeterministic high values and fixed low inputs. As SAT-based model checking is precise down to the individual bit, it will find a low input which triggers the underflow and uncovers the leak.

CBMC generates a counterexample falsifying a policy of e.g. no leakage and thereby having triggered the integer underflow. The following excerpt of the counterexample
\begin{small}
\begin{verbatim}
State 14 file underflow.c line 40 function main thread 0
----------------------------------------------------
  underflow::main::1::l=1706688912 (00000000...
....
State 35 file underflow.c line 13 function underflow thread 0
----------------------------------------------------
  underflow::underflow::1::nbytes=4027596816 (11110000...
\end{verbatim}
\end{small}
shows that a low input of \texttt{l=1706688912} lead to an \texttt{nbytes} which underflowed from the previous value 20.

Clearly, for such leaks to be detected it needs bit-level precise reasoning, just like SAT-based bounded model checkers support.
\subsection{Environment}

In model checking, the environment, like library function calls or generally functions and data structures which have no implementation, need to be modelled in a way which allows for the property to be verified. Out of the box, {\sc CBMC} replaces function calls with no implementation with nondeterministic values.

As our analysis needs to check for equality on inputs and outputs of functions a certain number of common library function have to be modelled in a way which preserves their original semantics. For example, the usual library C functions \texttt{memcmp}, and \texttt{strcmp} are implemented in a way which return 0 if their arguments are equal and a value not equal to 0 if they are not equal. The functions \texttt{memset} and \texttt{memcpy} actually set an array of integers or characters to a certain value or to the content of another array. The same applies to linux kernel utility functions such as \texttt{copy\_to\_user} and \texttt{copy\_from\_user} which copy memory blocks to or from userspace.

For example, a \texttt{memcmp} implementation is shown in Program \ref{PROG_3}.
\begin{program}
\begin{small}
\begin{verbatim}
int memcmp(char *s1, char *s2, unsigned int n) {
  int i;
  for(i=0;i<n;i++) {
     if(s1[i] != s2[i]) return -1;
  }
  return 0;
}
\end{verbatim}
\end{small}
\caption{Simplified \texttt{memcmp} model}
\label{PROG_3}
\end{program}

\section{Experimental Results}
We applied our technique to CVE reported information leakage vulnerabilities in the Linux Kernel. In the experiments we checked for policy violations and proved whether official patches resolve the information leakage. We also analysed authentication routines of the Secure Remote Password protocol (SRP) and of a Internet Message Support Protocol implementation. A summary of the results is shown in Table \ref{TBL_RESULTS}.
\begin{table}[t]
\centering
\begin{small}
\begin{tabular}{ l c c c c c r }
Description & CVE Bulletin & LOC & $k^{\star}$ & Patch Proof & $log_{2}(N)$ & Time\\
\hline \\
AppleTalk & CVE-2009-3002 & 237 & 64 &  \checkmark & 6 bit & 1h39m\\
tcf\_fill\_node & CVE-2009-3612 & 146 & 64 & \checkmark & 6 bit & 3m34s \\
sigaltstack & CVE-2009-2847 & 199 & 128 &  \checkmark & 7 bit & 49m50s \\
cpuset${}^{\dagger}$ & CVE-2007-2875 & 63 & 64 &  $\times$ & 6 bit & 1m32s  \\
\hline \\
SRP getpass & -- & 93 & 8 &  \checkmark & 1 bit & 0.128s\\
login\_unix & -- & 128 & 8 & -- & 2 bit & 8.364s\\

\hline

\end{tabular}
\end{small}
\caption{Experimental Results. $\star$ Number of unwindings $\dagger$ From Section \ref{SEC:MODELLING}}
\label{TBL_RESULTS}
\end{table}


\subsection{Linux Kernel}

We define information leakage in the kernel always as parts of the kernel memory which gets mistakenly copied to user space, i.e. the virtual memory allocated to conventional applications. Clearly, this should not happen as anything allocated in the kernel space is not meant to be seen by users (except within the bounds of normal user/kernel interactions), especially in multi-user systems like Linux. Thus, in all examples the kernel memory is modelled as nondeterministic values.

The interface between user and kernel space are system calls or syscalls in short. Syscalls, like normal functions, have a number of arguments and a return value where the kernel can transfer data structures or single values back and forth. This is the crucial point in the system where information leakage is most common.
\medskip

{\bf AppleTalk. } The specific vulnerability CVE-2009-3002 in the appletalk network code shows a quite common cause of information leakage: a user requests, by a syscall, that a structure gets filled with values and returned to user land. The developer however forgot to assign values to all fields in the struct, thus these missing fields get ``filled'' with unspecified kernel memory, as it is allocated on the stack. This CVE security bulletin actually comprises six different vulnerable network protocol implementations, all following the same leakage pattern. We will only present the affected code of the AppleTalk implementation -- the same kind of analysis applies to all six vulnerabilities.

In this case the structure returned to the user is shown in Program \ref{PROG_SOCK}.
\begin{program}
\begin{small}
\begin{verbatim}
struct sockaddr_at {
    u_char sat_len, sat_family, sat_port;
    struct at_addr      sat_addr;
    union {
        struct netrange r_netrange;
        char            r_zero[ 8 ];  
    } sat_range;
};
#define sat_zero sat_range.r_zero
\end{verbatim}
\end{small}
\caption{Complex observation struct leads to leak from \texttt{sat\_zero}.}
\label{PROG_SOCK}
\end{program}
The leaking function is \texttt{atalk\_getname} in \texttt{net/appletalk/ddp.c} is shown in Program \ref{PROG_ATALK}.

\begin{program}
\begin{small}
\begin{alltt}
static int atalk_getname(struct socket *sock, struct sockaddr *uaddr,
                         int *uaddr_len, int peer) \{
        struct sockaddr_at sat;
    
        // Official Patch. Uncommented to trigger leak
        //memset(&sat.sat_zero, 0, sizeof(sat.sat_zero));
        \vdots // sat structure gets filled
        memcpy(uaddr, &sat, sizeof(sat));
        return 0;
\}
\end{alltt} \end{small}
\caption{Function introducing the leak for CVE-2009-3002.}
\label{PROG_ATALK}
\end{program}
In the function, the structure \texttt{sat} gets filled with values provided by the kernel, at the end the whole structure is copied via \texttt{memcpy} to the address of the \texttt{uaddr} pointer, which is indirectly, via the syscall \texttt{getsockname} copied back to user land. However, the field \texttt{sat.sat\_zero} has not been initialised, thus a number of bytes of kernel memory get copied back to the user.

The secret is implicitly modelled by allocating the \texttt{sat} structure with nondeterministic values; observations are also of type \texttt{sockaddr\_at}. The driver uses as parameter \texttt{eq\_t} the library function \texttt{memcmp} to compare memories.

The model checker found a counterexample for a 6 bit policy within 1 hour and 39 minutes. Once the official patch was applied of setting the \texttt{sat} structure to 0 with \texttt{memset}, our driver successfully verified the policy in about the same time with unwinding assertions, thus it proved that the patch stops the leak.
\medskip

{\bf tcf\_fill\_node. } This information leak happens in the netlink subsystem of the kernel. The function \texttt{tcf\_fill\_node}  prepares a \texttt{struct tcmsg} to be sent back to the user. However, the programmer made a typing mistake and filled a field \texttt{tcm\_\_pad1} twice instead of the second time for \texttt{tcm\_\_pad2}.

\begin{program}
\begin{small}
\begin{alltt}
struct tcmsg *tcm;
...
nlh = NLMSG_NEW(skb, pid, seq, event, sizeof(*tcm), flags);
tcm = NLMSG_DATA(nlh);
tcm->tcm_family = AF_UNSPEC;
tcm->tcm__pad1 = 0;
tcm->tcm__pad1 = 0; // typo, should be tcm__pad2 instead.
\end{alltt}
\end{small}
\caption{Function excerpt introducing the leak for CVE-2009-3612.}
\end{program}
This leaks kernel memory from \texttt{tcm\_\_pad2} back to userspace. Here, we again modelled kernel memory implicitly by the memory allocated for \texttt{tcm} through the function \texttt{NLMSG\_DATA}, which initialised the fields of the struct with nondeterministic values. The observation is the filled out variable \texttt{tcm}, the low user input is a simple integer variable not mentioned here for brevity.

The official patch which was applied to fix the leak is simply changing the last line above to \texttt{tcm->tcm\_\_pad2=0}. We were again able to prove that this patch successfully fixes the security hole and otherwise the program violates a leakage policy of 6 bits.

Without the patch, a counterexample is found within 3 minutes and 34 seconds; with the patch, the program is verified within about the same time.
\medskip

{\bf sigaltstack. } The leakage for this vulnerability is intricate and only manifests itself on 64-bit processors. On such a system, the struct \texttt{stack\_t}, as shown in Program \ref{PROG_SIGNAL}, will be padded to a multiple of 8 bytes because on 64-bit systems \texttt{void*} and \texttt{size\_t} are both 8 bytes (instead of 4 bytes for 32-bit systems), while an integer type remains 4 bytes. Thus, the size of \texttt{stack\_t} is padded to 24 bytes, while on a 32-bit system it remains unpadded at 12 bytes.

\begin{program}
\begin{small}
\begin{verbatim}
typedef struct sigaltstack {
        void __user *ss_sp;
        int ss_flags; // 4 bytes padding on 64-bit systems
        size_t ss_size;
} stack_t;
\end{verbatim}
\end{small}
\caption{Structure with padding depending on architecture.}
\label{PROG_SIGNAL}
\end{program}
The syscall \texttt{do\_sigaltstack} in \texttt{kernel/signal.c} copies such a structure back to userland via the copy function \texttt{copy\_to\_user}, however it does not clear the padding bytes, thus those are leaked to the user on a 64-bit system. In the function visible in Program \ref{PROG_SIGNAL2}, the high input is the structure \texttt{oss} and the low output is the argument \texttt{uoss}.

\begin{program}
\begin{small}
\begin{verbatim}
int do_sigaltstack (const stack_t __user *uss, stack_t __user *uoss, 
                    unsigned long sp) {
     stack_t oss;
      ... // oss fields get filled
      if (copy_to_user(uoss, &oss, sizeof(oss)))
             goto out; ....
\end{verbatim}
\end{small}
\caption{Leakage through copying whole structures including padding.}
\label{PROG_SIGNAL2}
\end{program}

CBMC supports modelling of 64-bit widths however that is not enough to automatically measure the padding bytes. This is because the \texttt{sizeof} operator in CBMC returns only the sum of all sizes without eventual bit alignments. This is solved in our approach by providing a model of the \texttt{copy\_to\_user} function, just like e.g. an implementation of \texttt{memcpy} is provided, which checks if the length parameter is aligned according to the architecture (4 bytes for 32 and 8 bytes for 64). If there are padding alignments then these will be chosen to be filled with nondeterministic integer values modulo the number of padding bytes.

In Program \ref{PROG_SIGNAL2}, this would translate to the following: \texttt{sizeof(oss)} counts 20 bytes as the size of the structure. However, this does not account for the padding bytes, and our \texttt{copy\_to\_user} model does the following calculation:
\begin{quote}
\begin{small}
\begin{alltt}
pad = ALIGN - (sizeof(oss) % ALIGN);
if(pad == ALIGN) padding = 0;
else padding = ((unsigned int) nondet_int()) % (1 << (pad*8))
\end{alltt}
\end{small}
\end{quote}
where \texttt{ALIGN} is chosen to be 4 or 8 depending on the architecture used. In a 64-bit system, this translates to $8 - (20 \% 8) = 4$ bytes for \texttt{pad} which are represented by the \texttt{padding} variable.

With this setup, we were able to verify that on a 32-bit system the Program \ref{PROG_SIGNAL2} does not leak anything, while on a 64-bit system this violates a policy of e.g. 7 bits. A counterexample was found within 49 minutes and 50 seconds. We were also able to prove that the patch applied removes the padding leak. The patch in this case was to not copy the whole struct but copying the three struct members separately through the function \texttt{\_\_put\_user}, where the padding does not come into play.
\medskip

{\bf cpuset. } The crucial part of this vulnerability has already been discussed in Section \ref{SEC:MODELLING}. Our analysis finds the right low input which triggers the integer underflow. The actual code however does not simply return the secret as shown in the section mentioned above, but it copies \texttt{nbytes} number of bytes from a buffer \texttt{ctr->buf} at offset \texttt{*ppos}.
\begin{program}
\begin{small}
\begin{verbatim}
if (*ppos + nbytes > ctr->bufsz)
      nbytes = ctr->bufsz - *ppos;
if (copy_to_user(buf, ctr->buf + *ppos, nbytes))
     return -EFAULT;
\end{verbatim}
\end{small}
\end{program}
Because of the underflow, \texttt{nbytes} and \texttt{*ppos} access memory way out of the actual buffer and thus disclose kernel memory. However our analysis of this vulnerability requires at the moment too much manual intervention to model memory access outside of the allowed bound ( i.e. \texttt{ctr->buf + *ppos}).

One elegant way of addressing this  problem would be by modifying CBMC itself; CBMC could for example return nondeterministic values for such out-of-bound memory accesses which would implicitly model the access to confidential data.

\subsection{Authentication Checks} We analysed parts of the authentication routines of the secure remote password suite (SRP) and the Unix passwd-authentication of Cyrus' Internet Message Support Protocol daemon (IMSPD).

{\bf SRP. } To demonstrate that confidential variables and observations can be used flexibly, we checked that there is no leakage in the password request function in \texttt{libsrp/t\_getpass.c}.

The confidential input is the password entered by the user when being prompted at the login; the observations are the \textit{echos} of the terminal of typed characters. Whether the terminal echos the typed characters or not depends on which mode the console is in. The environment modelling the console and its modes had to be provided to check this program.

\begin{program}
\begin{small}
\begin{verbatim}
_TYPE( int ) t_getpass (char* buf, unsigned maxlen, 
                                   const char* prompt) {
    DWORD  mode;

    GetConsoleMode( handle, &mode );
    SetConsoleMode( handle, mode & ~ENABLE_ECHO_INPUT );
    if(fputs(prompt, stdout) == EOF ||
        fgets(buf, maxlen, stdin) == NULL) {
        SetConsoleMode(handle,mode);
        return -1;
    } ....
\end{verbatim}
\end{small}
\caption{Side-effect of \texttt{mode} decides on echo output of \texttt{fgets}}
\label{PROG_5}
\end{program}

In Program \ref{PROG_5}, the function \texttt{t\_getpass} first gets the current mode of the console by the function \texttt{GetConsoleMode}; then it sets a new console mode by inverting the bit \texttt{ENABLE\_ECHO\_INPUT} in the mode through the function \texttt{SetConsoleMode} which clearly disables the echo of input read from standard input. The function \texttt{GetConsoleMode} is modelled by nondeterministically setting the mode to any integer value, the function \texttt{SetConsoleMode} sets a global mode variable to its second argument. The function \texttt{fgets}, which reads a number of bytes from \texttt{stdin}, is modelled to return its first argument \texttt{buf} completely if the mode is set to echo the input and return a constant value otherwise.

With this setup CBMC proves through our driver that starting from any initial \texttt{mode}, the program will always end up with $\log_2(| \simeq_P |)=0$, i.e. that there is no leakage. We can also successfully check that if the line which disables the echo is removed then the policy is violated.

{\bf IMSPD. } The function checked in this test is \texttt{login\_plaintext} in \texttt{imsp/login\_unix.c} as shown in Program \ref{PROG_LOGIN}.

\begin{program}
\begin{small}
\begin{verbatim}
int login_plaintext(char *user, char* pass, char** reply) {
    ...
    struct passwd* pwd = getpwnam(user);
    if (!pwd) return 1;
    if (strcmp(pwd->pw_passwd, crypt(pass, pwd->pw_passwd)) != 0) {
        *reply = "wrong password";
        return 1;
    }
    return 0;
\end{verbatim}
\end{small}
\caption{Login function of IMSPD.}
\label{PROG_LOGIN}
\end{program}
\noindent The program first tries to receive the stored password context of a user using the function \texttt{getpwnam}. If successful, it will compare the stored with the entered password using \texttt{strcmp}. If this fails it will set the string \texttt{reply} to ``wrong password''. If authentication is successful it returns 0.

Clearly, this function has three distinguishable observations: (1) it returns 1 (2) it returns 1 and sets \texttt{*reply} (3) it returns 0. We modelled the three parameters to the function as low user input and the stored password as confidential variable. With this setup, we are able to verify that this program conforms to a policy which only leaks 3 observations, within 9 seconds.

\section{Related Work}
There have been several attempts in recent years to build a quantitative analysis of leakage, starting with the static analysis in
\cite{CHM05}.

The most relevant works for this paper  are \cite{disquant} by M. Backes, B. K{\"o}pf and A. Rybalchenko and \cite{aqua} by J. Heusser and P. Malacaria where verification techniques are used to compute leakage of programs.
Those works are both inspired by the important previous theoretical work on self composition by G. Barthe, P. D'Argenio,  and T. Rezk \cite{rezk} and T. Terauchi and A. Aiken \cite{safety}. 
However as already noted, those approaches attempts primarily to answer questions about how much a program leak and seem unable to scale to real code in terms of line of code, state space and language constructs. In particular, they have not, as far as we are aware, been used to analyse independently existing vulnerabilities in independently existing programs.

 On the theoretical side, the complexity of QIF analysis has recently been thoroughly investigated by H. Yasuoka and T. Terauchi \cite{YaTe10} who, amongst other aspects, explored the relation to verification and k-safety properties.

Approaches that do scale to large programs are by S. McCamant, M. D. Ernst \cite{McCamant2008} and J. Newsome, S. McCamant, D. Song \cite{McCamant09}. They released an impressive tool, FlowCheck, which is able to analyse very large programs. There are however significant differences between the approaches in that FlowCheck is a security testing tool based on the Valgrind dynamic instrumentation framework whereas our approach is based on verification and static analysis techniques. Thus, our work comes with stronger theoretical guarantees (for example verification of the official patches) and does not require to ``run'' the code.

D. Kroening's CBMC \cite{ckl2004} has been used for many practical applications. A good overview over the applied fields can be found under the following link \cite{cprover}.

\section{Conclusion}

In this paper we combined state of the art model checking with theoretical work on Quantitative Information Flow, to provide a powerful tool for the analysis of leakage of information. We demonstrated not only that CVE reported vulnerabilities such as for the Linux kernel can be analysed with a level of scalability and precision able to find real security vulnerabilities,
but that it is also possible to prove whether the official patches fix the problem. 
We argued that leaks are not synonymous of a security breach and hence a quantitative framework is better equipped than a qualitative one to determine when an information leak represents a security threat. 

We see this work as a significant step in the application of academic research on information flow analysis to real-world problems in systems software.

\end{document}